# Overcoming Thermo-Optical Dynamics in Broadband Nanophotonic Sensing

Mingkang Wang[1,2,†], Diego J. Perez-Morelo[1,2], Vladimir Aksyuk[1,*]

[1]Microsystems and Nanotechnology Division, National Institute of Standards and Technology, Gaithersburg, MD 20899 USA

[2]Institute for Research in Electronics and Applied Physics, University of Maryland, College Park, MD 20742, USA

Correspondence:

† Mingkang Wang, tel:+1-301-605-4531; mingkang.wang@nist.gov

*Vladimir Aksyuk, tel: +1-301-975-2867; vladimir.aksyuk@nist.gov

## Abstract

*Advances in integrated photonics open exciting opportunities for batch-fabricated optical sensors using high quality factor nanophotonic cavities to achieve ultra-high sensitivities and bandwidths. The sensitivity improves with higher optical power, however, localized absorption and heating within a micrometer-scale mode volume prominently distorts the cavity resonances and strongly couples the sensor response to thermal dynamics, limiting the sensitivity and hindering the measurement of broadband time-dependent signals. Here, we derive a frequency-dependent photonic sensor transfer function that accounts for thermo-optical dynamics and quantitatively describes the measured broadband optomechanical signal from an integrated photonic atomic-force-microscopy nanomechanical probe. Using this transfer function, the probe can be operated in the high optical power, strongly thermo-optically nonlinear regime, reaching a sensitivity of $\approx 0.4$ fm/Hz$^{1/2}$, an improvement of $\approx 10\times$ relative to the best performance in the linear regime. Counterintuitively, we discover that higher transduction gain and sensitivity are obtained with lower quality factor optical modes for low signal frequencies. Not limited to optomechanical transducers, the derived transfer function is generally valid for describing small-signal dynamic response of a broad range of technologically important photonic sensors subject to the thermo-optical effect.*

## Introduction

The rapid development in integrated photonics and nanotechnology has enabled a wide range of nanophotonic sensors, applicable for thermal [1,2], magnetic [3,4], gas [5,6], force [7] and displacement [8,9] sensing. High quality factor (Q) nanophotonic cavities strongly enhance local light-matter interaction via their small optical mode volumes and extended photon lifetimes, enabling an unmatched combination of ultrahigh precision and ultrawide bandwidth for optical sensing in a variety of applications. This includes the cavity-optomechanical on-chip motion transduction[10,11], where for example, a low loss nanomechanical resonator has been combined with a high finesse optical cavity for ultrafast nanoscale optomechanical atomic force microscopy (AFM) [12–14], and recently applied for direct measurement of local chemical and thermal properties using photothermal-induced resonance (PTIR) [15].

Generally, signal-to-noise of nanophotonic sensors improves with higher transduction gains obtained by increasing the optical power in the photonic cavity. However, the very same beneficial

qualities – small volumes and high Q – increase local optical heating, which quickly becomes significant at even modest input powers. This limits the power that can be used for better sensitivity since with further power increase the cavity frequency becomes power- and time-dependent due to the thermo-optical effects[16–19] complicating the transduction dynamics and making it difficult to recover the sensor input stimulus signal from the detected photonic response for any broadband stimuli. Taking the AFM application as an example, simultaneous and accurate measurements of static as well as dynamic probe displacements are critical to obtaining long-range and short-range sample topography during rapid scanning. It is also key to measuring and quantitatively interpreting dynamic processes at the nanoscale[15,20]. An accurate quantitative description of the optical-power-dependent sensor transfer function (the change of the photonic response per unit mechanical displacement) across the mechanical frequency range down to DC will remove the present input power limitation and push the sensitivity to the next level. Not limited to optomechanics, the enhanced thermo-optic effect is ubiquitous in nanophotonic cavity sensing, and a transfer function accounting for it would broadly enable operating this class of sensors with increased sensitivity and transduction bandwidth.

The present work aims to advance nanophotonic sensing, leveraging the highly integrated optomechanical AFM probes [14,15] as a well established high-performance experimental platform. We derive and experimentally confirm an optical power- and frequency-dependent optomechanical transfer function working down to the low-frequency range where at high optical power the optomechanical transduction gain is affected by the thermo-optical tuning. We first use the thermodynamic fluctuations of a nanomechanical cantilever as a broadband stimulus, measured by an evanescently-coupled microdisk in the thermo-optically nonlinear regime. The derived transfer function, requiring no adjustable parameters, is found to accurately describe the photonically measured signal power spectral density over a wide range of frequencies and different degrees of thermo-optical nonlinearity. Additionally, by driving the cantilever and calibrating the displacement, we demonstrate and quantify the increased transduction gain and sensitivity in the high optical power regime. When the AFM probe operates in the thermo-optical nonlinear regime, the sensitivity is increased by $\approx$ 10 times compared to the best performance in the linear regime, reaching $\approx 0.7$ fm/Hz$^{-1/2}$. Furthermore, by considering the effect of the optical mode Q on the transduction gain with thermal dynamics included, we discover and experimentally confirm that, counterintuitively, a lower Q cavity mode could provide a larger transduction gain, and sensitivity, for low-frequency signals, compared to a higher Q mode. Understanding signal transduction in the presence of thermo-optical dynamics enables the high-power, high-sensitivity operation of a general class of photonic sensors, with the optomechanical AFM probe being just one example.

## Results and discussion

### Experimental setup

The nanophotonic sensor under study is an optomechanical device consisting of a curved cantilever probe held in the nearfield of a microdisk optical cavity supporting whispering-gallery modes (WGMs). The experimental demonstration of the full operation of this device as AFM sensors has been reported[14], and it has enabled unprecedented new data collection regime in AFM/PTIR setup. Fig. 1(a) shows an electron micrograph of the optomechanical AFM probe. The cantilever has multiple mechanical modes, however, only the fundamental in-plane mode of a

(5.46 ±0.01) MHz eigenfrequency is dominant, while other mechanical modes (for both the cantilever and the microdisk itself) are either far above the fundamental mode or do not produce a strong optical readout signal due to motion being mostly orthogonal to the microdisk plane. In some experiments, we cleaved the device such that the cantilever was supported at only one end (cleave at the red dashed line in the inset), lowering the mechanical stiffness and the fundamental in-plane mechanical mode frequency to (310 ± 0.2) kHz, which is beneficial for studying the frequency-dependent transfer function in the thermally-driven case. Note, the microdisk thermo-optical nonlinearity effects are the same for both the double-clamped and single-clamped cantilever probes. Throughout the paper, the reported uncertainties are one standard deviation statistical uncertainty, unless noted otherwise.

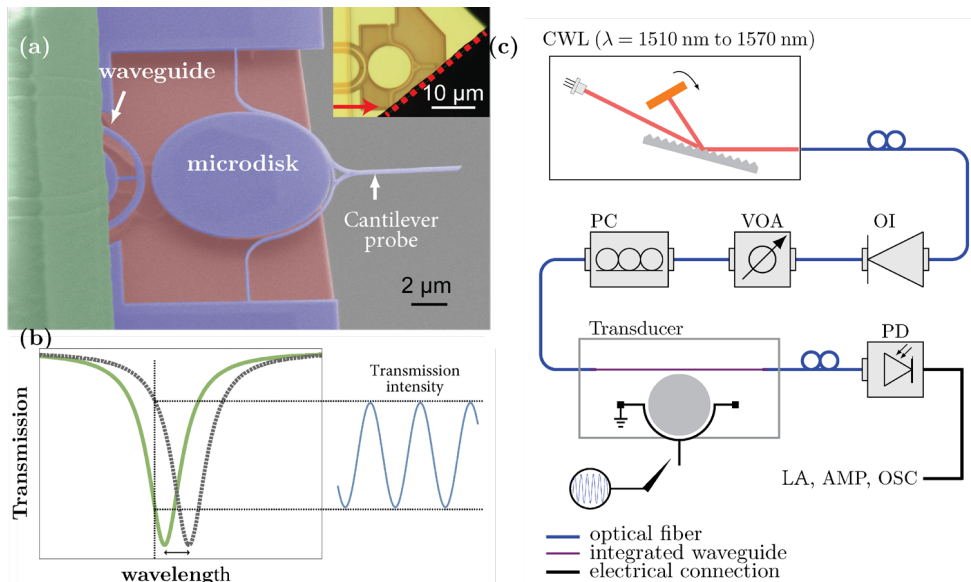

Figure 1 Introduction of the system (a) Colorized scanning electron micrograph of an optomechanical AFM probe with a double-clamped cantilever. The cantilever has a nominal width of 150 nm and is separated by a gap of 200 nm from a 10 μm diameter silicon microdisk. The cantilever and the silicon disk are nominally 260 nm thick. The disk is supported by a silica post underneath. An optical photo of a single-clamped device is shown in the inset. The red arrows and the dashed lines label the position where the cantilever is cleaved from the anchor point. (b) During the measurement, the fiber-coupled laser wavelength is fixed (vertical dashed line). The motion of the cantilever changes the cantilever-disk gap, shifting the resonances of the microdisk. At the working wavelength on the shoulder of the optical resonance, the shifting resonance generates a varying transmission intensity. (c) Schematic of the measurement setup containing CWL (continuous wavelength tunable diode laser), optical isolator (OI), variable optical attenuator (VOA), polarization controller (PC), electronic amplifier (AMP), lock-in amplifier (LA), oscilloscope (OSC), photodetector (PD). The cantilever is grounded, and a metal tip drives the cantilever by applying an electrostatic force.

The integrated optical cavity works as a high Q interferometric motion transducer where the motion of the cantilever modulates the optical mode frequencies. The input laser is tuned to a shoulder of the cavity resonance, such that at small optical powers where the thermo-optical effect is negligible, the variations in the transmitted optical power are proportional to the displacement of the cantilever, as shown in Figure 1(b).

Figure 1(c) presents the setup for measuring the optical and mechanical responses of the system. Transmission measurements use a continuous-wavelength tunable laser (1510 nm to 1570 nm) as the light source. Laser light is injected from a single-mode fiber into an integrated waveguide, which is evanescently coupled to the microdisk cavity of the optomechanical AFM probe. The transmitted light is coupled from the waveguide output into an output fiber and detected by a photodetector connected to a preamplifier, lock-in amplifier, and oscilloscope. The device is measured in air at room temperature. The cantilever is driven by the stochastic Langevin force from the thermal bath and an electrostatic force applied from a metal tip in close proximity to the cantilever.

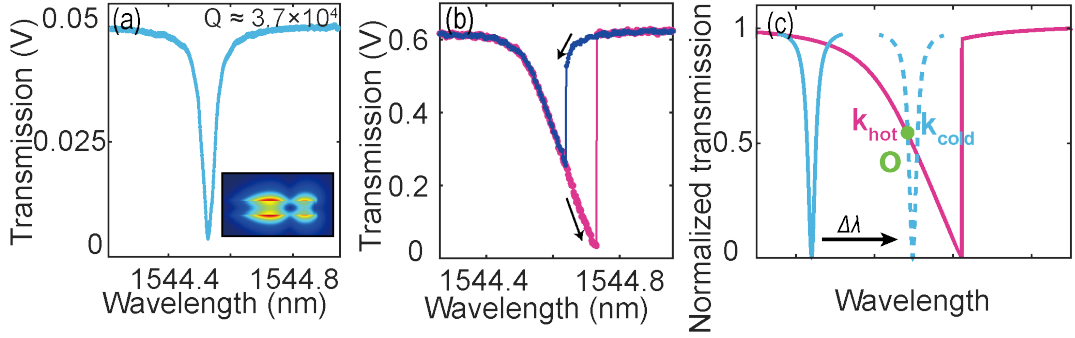

*Figure 2* Whispering-gallery mode and thermo-optical nonlinearity. Linear and thermo-optically nonlinear transmission spectra of a WGM are shown in (a) and (b), for low and high input optical power, respectively. The radial-cross-sectional mode-shape, i.e. the distribution of electric field intensity, is shown in the inset. The hysteresis in (b) is obtained by changing the direction of the laser wavelength sweep, as labeled by the black arrows. Q $\approx 3.7 \times 10^4$. (c) Schematic of the thermo-optical nonlinearity. The thermally-induced refractive index change tunes the resonance frequency during wavelength sweep (from solid blue line to dashed blue line), making it intensity-dependent, shown as the purple line. The absolute value of the slopes for the normalized linear and nonlinear spectra at a working wavelength O are $k_{cold}$ and $k_{hot}$, which are proportional to their corresponding transduction gains.

Figures 2(a) and 2(b) show a WGM resonance dip in the transmission spectrum of the device under low and high input powers, respectively. With increasing the light intensity, the optical response exhibits the nonlinear effect evident from the hysteresis shown in Fig. 2(b). It is noteworthy that the slope of the high-power transmission dip is linear, suggesting that the resonance spectral shift from the thermo-optical tunning is linearly proportional to the power dropped into the microdisk, as discussed in detail later.

The broadening, asymmetry, and hysteresis in the transmission dip in such small Si cavities are generally attributed to the linear and nonlinear two-photon optical absorption in conjunction with the thermo-optical effect[16], rather than the optical nonlinearity[21]. The small mode volumes and high Q of the cavities enhance the light-matter interaction, enabling sensitive measurement of the high-mechanical-bandwidth nanoscale cantilever probes, but also increase the local cavity heating and thermo-optical tuning. Miniaturization of the mechanical probes extends their useful mechanical transduction bandwidth from DC up to at least 10 MHz, a range that includes timescales both well below and high above the typical few-microsecond thermal time constants of nanophotonic cavities[22], therefore requiring accurate quantification of thermal effects on signal transduction across the frequency range.

**Transfer function in the thermo-optically nonlinear regime**

Temperature changes shift the cavity resonances via two main mechanisms, through the thermal expansion of the disk and the refractive index change of the cavity materials. For a small change in temperature $\Delta T$, the relative wavelength shift can be written as:

$$\frac{\Delta \lambda}{\lambda_0} = \left(\chi + \frac{1}{n_0}\frac{dn}{dT}\right)\Delta T \quad (1)$$

where $\chi = 2.6 \times 10^{-6}$ K$^{-1}$ is the thermal expansion coefficient of silicon, $n_0 \approx 3.48$ is the refractive index of silicon at the working wavelength $\lambda_0 \approx 1.5$ μm at room temperature, $\frac{dn}{dT} = 1.72 \times 10^{-4}$ K$^{-1}$ is the temperature sensitivity of the refractive index[23] and $\Delta T$ is the temperature change induced by the absorbed optical power. Eq. (1) can be rewritten as $\frac{\Delta\lambda}{\lambda_0} = \alpha \Delta T$, where specifically for Si $\alpha \approx \frac{1}{n_0}\frac{dn}{dT} = 4.94 \times 10^{-5}$ K$^{-1} \gg \chi$. During the sweep of the laser wavelength across a WGM resonance the cavity temperature increases linearly with the absorbed power. As shown in Figure 2(c), the rising temperature continuously shifts the optical resonance (blue dashed line) to longer wavelengths, generating the nonlinear optical transmission dip in the spectrum (purple line). At a working point $O$ on the spectrum, we define the absolute value of the slope of the normalized "hot cavity" transmission dip as $k_{hot}$ and the considerably steeper slope of the thermally-shifted linear optical response (cold cavity) as $k_{cold}$.

Typical photonic cavities have short photon lifetimes, responding near-instantaneously to mechanical motion in the absence of heating effects. The thermo-optical effect complicates the optomechanical transfer function, coupling it to the thermal dynamics of the nanophotonic cavity. For a given microdisk incident intensity $I_i$ (background intensity at off-resonance wavelengths), and a transmitted intensity $I < I_i$ at $O$, the small transmitted intensity change $\Delta I$ near the working point is a linear function of both the displacement $\Delta x$ and the temperature change $\Delta T$:

$$\frac{\Delta I[x, \Delta T(I)]}{I_i} = k_{cold}(g_{om,\lambda}\Delta x + \lambda_0 \alpha \Delta T) \quad (2)$$

where $g_{om,\lambda} = \frac{d\lambda}{dx}$ is the optomechanical coupling coefficient for the cavity wavelength. In turn, the temperature $T$ is a function of the intensity lost into the cavity $I_i - I$. The small temperature variation $\Delta T$ near $O$ can be obtained by considering the thermodynamic equation for the optical cavity with a thermal time constant $\tau$:

$$\frac{d(\Delta T)}{dt} = -\frac{1}{\tau}(\Delta T + \eta \Delta I) \quad (3)$$

where $\eta$ is the proportionality constant between the temperature change $\Delta T$ and the transmitted intensity change $\Delta I$, whereby decreased transmitted intensity corresponds to increased heating. $\eta$ is proportional to the product of the thermal impedance of the sensor, and the ratio of the energy absorption to the total energy loss in the cavity (which also includes radiative leakage and scattering loss).

The complete frequency-dependent transfer function of the linearized system (2),(3) can be obtained by considering the system's response at the working point $O$ to a small harmonic cantilever motion, which perturbs the cavity resonance via the optomechanical coupling. The optomechanically induced periodic resonance frequency shift modulates the power entering the cavity, resulting in both a small harmonic temperature variation $\Delta T$ as well as the transmitter intensity variation $\Delta I$:

$$\Delta x = x_0 e^{i\omega t}$$
$$\Delta T = T_0 e^{i\omega t} \quad (4)$$
$$\Delta I/I_i = i_0 e^{i\omega t}$$

where the harmonic displacement of the cantilever has an amplitude of $x_0$ and frequency of $\omega$, $T_0$ is the amplitude of harmonic temperature variation and $i_0$ is the relative intensity modulation amplitude. By combining Eq. (4) with Eq. (2), (3), we obtain the small-signal thermal, $T_0$, and intensity modulation, $i_0$, responses as:

$$T_0 = -\eta I_i \frac{r}{i\omega \tau r + 1} \beta x_0 \quad (5)$$

$$i_0 = \Lambda(\omega) x_0 \quad (6)$$

where $r$ is defined as $r = \frac{1}{1+k_{cold}\lambda_0 \alpha \eta I_i}$, $\beta = k_{cold} g_{om,\lambda}$, and $\Lambda(\omega) = \left(1 - \frac{1-r}{i\omega\tau r+1}\right)\beta$ is the complete frequency-dependent transfer function connecting input signal amplitude $x_0$ with the resulting relative intensity modulation $i_0$. When either the incident intensity $I_i$ is low or the intensity change is not converted into a temperature change ($\eta \approx 0$), we have $r \approx 1$, the microdisk temperature change $\Delta T \approx 0$, and $\Delta I/I_i = \beta \Delta x$ is linear, with a constant transfer coefficient $\beta$. With increasing the incident intensity $I_i$, the ratio $r$ decreases below unity and the transfer function acquires the correction coefficient $\left(1 - \frac{1-r}{i\omega\tau r+1}\right) < 1$ arising from thermal dynamics and the thermo-optical effect. The magnitude of the transmission modulation is

$$|i_0| = |\Lambda| x_0 = \beta x_0 r \sqrt{\frac{1+\tau^2 \omega^2}{1+\tau^2 \omega^2 r^2}} \quad (7)$$

Physically, the transfer function is equivalent to a high-pass filter, which attenuates the transduction gain of the sensor in the adiabatic frequency limit, $\omega \ll 1/\tau$, by a factor of $r$: $|i_0| = r\beta x_0$. At high frequencies $\omega > 1/\tau$ the thermal dynamics does not play a role, $|i_0| = \beta x_0$, since the temperature change cannot follow the quick intensity modulation. By comparing the transfer function at the adiabatic limit and high-frequency limit, we see that the attenuation factor $r$ describes the ratio of hot and cold cavity slopes, $r = \frac{k_{hot}}{k_{cold}}$. Experimentally, $r$ is obtained directly from the ratio of the slopes of the measured normalized transmission spectra [Fig. 3(a)]. The thermal time constant $\tau$ can be characterized experimentally by a number of common methods, including from time-domain or frequency-domain response to a stimulus (e.g. optical power or frequency change), or via solving the heat equations via a finite element method.

Fig. 3(a) shows the normalized transmission signal of the optical mode. The scanning speed of the laser is $\approx 6$ nm/s which guarantees the spectrum is taken in the adiabatic limit. With increasing the input optical powers $P_{in}$ ($\propto I_i$), the dip in the spectrum becomes increasingly asymmetric due to the thermo-optical nonlinearity, and the slope decrease from $k_{cold}$ to $k_{hot1}$, $k_{hot2}$ and $k_{hot3}$, from which we obtain the ratios $r_{1,2,3} = \frac{k_{hot1,2,3}}{k_{cold}}$.

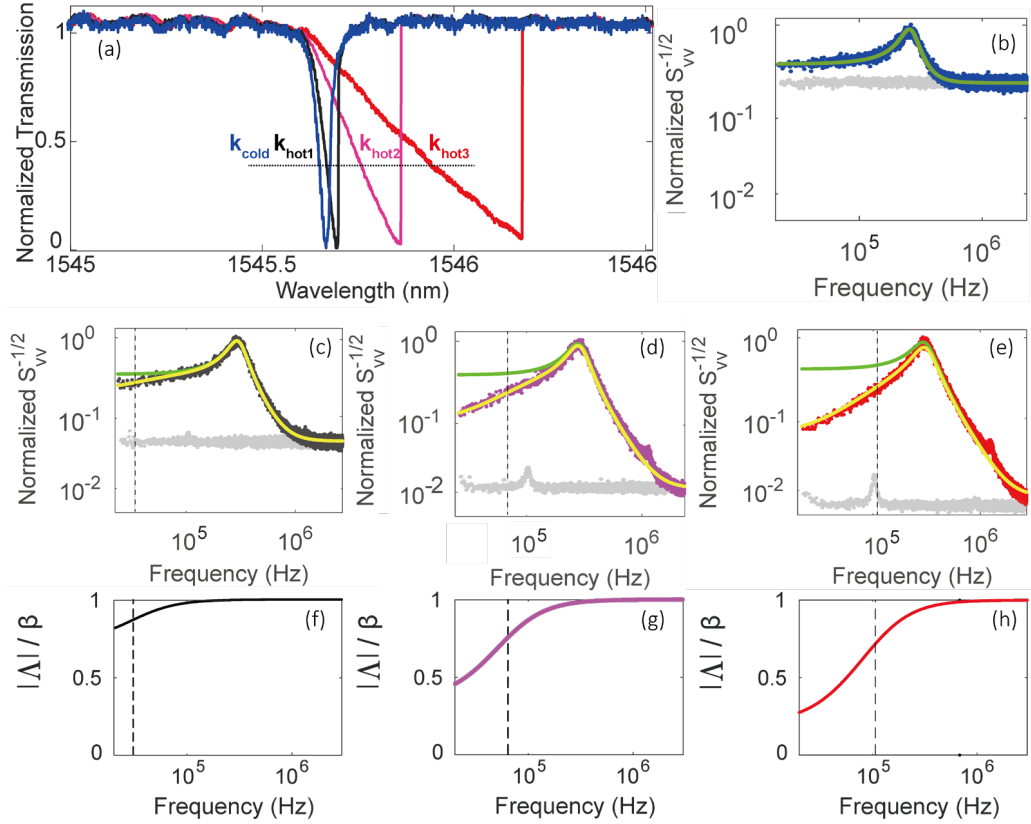

*Figure 3.* Optothermal-induced changes in the transfer function. (a) Normalized transmission signal with slopes of $k_{cold}$, $k_{hot1}$, $k_{hot2}$ and $k_{hot3}$. The cross points between the horizontal dashed line and the transmission signal mark the working wavelengths for the (b-e). (b-e) Normalized power spectral density of the optomechanically detected single-clamped cantilever's thermal-mechanical fluctuations. They correspond to the normalized transmission data of the same colors shown in (a). The green lines are the Lorentzian fit plus the measured background noise (gray dots), and the yellow lines are given by the nonlinear optical transfer function with no adjustable parameters needed. (f-h) show the transfer function in (c-e) normalized by $\beta$. The vertical dashed lines mark $1/(2\pi r\tau)$. $r_{1,2,3} = k_{hot1,2,3}/k_{cold} \approx 0.74, 0.36$ and $0.21$. The low-input-power case (b) corresponds to a linear transmission dip therefore it has the normalized transfer function of unity.

The thermally driven resonator is an ideal platform to demonstrate the transfer function since the Langevin force is frequency-independent ('white') and simultaneously excites equally at all frequencies. To ensure the mechanical fluctuation signal is well above the detection noise floor, we cleaved the device to obtain a softer, single-clamped cantilever. It lowers the mechanical stiffness to $\approx 32$ mN/m and the fundamental in-plane mechanical mode frequency to $\approx 310$ kHz.

Fig 3(b) presents the normalized measured power spectral density as a function of the frequency $f = \frac{\omega}{2\pi}$ for the linear optical mode shown as the blue dip in Fig. 3(a). The green line shows the corresponding Lorentzian fit (Q $\approx 3.0$ due to air damping) plus the independently measured detection noise shown as the gray dots. Figure 3 (c)-(e) present the normalized power spectral density as the laser intensity is increased, corresponding to the transmission spectra of the same color shown in Fig. 3(a). Larger spectral shifts are achieved ($1 > r_1 > r_2 > r_3$). The green lines are the corresponding Lorentzian fit performed at $f$ close to the peaks and larger than $1/(2\pi r\tau)$ (vertical dashed lines), where the thermo-optical-nonlinearity-induced attenuation is small. The separately measured detection noise (gray dots) is added to the Lorentzian fit. The peak on the background at around 0.1 MHz is attributed to the vibration of floating parts of the waveguide. Note, as the thermal response has the same magnitude in spite of different transduction gain, the normalized detection noise floor becomes lower with a larger transduction gain, meaning a better detection sensitivity [gray dots in Figure (c)-(e)]. At $f \gg 1/(2\pi r\tau)$, the measured power spectral density agrees well with the Lorentzian fit. The deviation appears at $f < 1/(2\pi r\tau)$ where the thermo-optical response becomes significant.

Fig. 3(f-h) show the corresponding transfer function magnitude $|\Lambda|$ normalized by $\beta$. This normalized optomechanical transduction coefficient is around $\frac{|\Lambda|}{\beta} \approx 1$ for $f \gg 1/(2\pi r\tau)$ and monotonically decreases with lower frequency, tending to $\frac{|\Lambda|}{\beta} \approx r$ at $f \ll 1/(2\pi\tau)$. Based on Eq. (7), the yellow lines in Fig. 3(c-e) are generated by multiplying the Lorentzian fits $\beta\, x_0(\omega)$ (green lines) by the normalized transfer function $|\Lambda|/\beta$. $r_{1,2,3}$ are obtained independently from the slopes in Fig. 3(a). For our microdisk cavity, we use $\tau \approx 7.0$ μs obtained independently by the finite element method numerical modeling [see Supplementary Materials]. Therefore, the transfer function does not include any adjustable parameters, while good agreement is achieved for all powers.

One consequence of thermo-optical tuning is that the transduction gain at low frequency will saturate with increasing input power. In the linear regime, the transduction gain is proportional to the input optical power $\propto P_{in} \propto I_i$, i.e. the same amount of resonance shift induces a larger transmitted power variation for a higher $I_i$. However, the transfer function is attenuated by a factor of $r$ at the adiabatic limit [$f \ll 1/(2\pi\tau)$] in the nonlinear regime, saturating the transduction gain $\propto rI_i$ with increasing input powers as $rI_i = \frac{I_i}{1+k_{cold}\lambda_0\alpha\eta I_i} \approx \frac{I_i}{k_{cold}\lambda_0\alpha\eta I_i} \propto \frac{I_i}{I_i}$ when $r \ll 1$ [see Supplementary Materials]. Note, the saturation only occurs in the low-frequency region, the linear dependence of the gain on $I_i$ at high frequency [$f \gg 1/(2\pi\tau)$] is still valid in the nonlinear regime.

**Improvement of displacement sensitivity by operating in the nonlinear regime**

The derived high-incident-power transfer function in the thermo-optically nonlinear regime is easily invertible and enables increased sensitivity for broadband measurements of probe displacement. To demonstrate the sensitivity increase, a probe response to an external drive force is measured in the high-power, nonlinear regime and compared to the best measurement in the linear regime. The measurement is performed on a regular probe with a double-clamped cantilever that we used in previous experiments[14,15] [electron micrograph in Fig. 1(a)]. The cantilever is driven by an electrostatic force applied via a metal tip in close proximity to the cantilever [Fig. 1(c)]. The optical transmission carrying the displacement signal is demodulated at the frequency of the drive by a lock-in amplifier.

Figure 4(a) shows the best achievable performance of the AFM probe in the linear regime where the input optical power is set at the threshold where the nonlinearity is about to appear. Figure 4(b) shows the data measured with the same drive strength but with increased optical power in the nonlinear regime, where the much stronger sensor signal is evident for the same driven motion. The electrostatically-driven response (purple dots) is measured while scanning the driving frequency. The measurement noise, shown as the red dots, is recorded while the drive is off, and consists of the probe's thermal-mechanical noise added to the detection noise background (gray dots). Measured noise voltage is plotted with a 1 Hz equivalent noise bandwidth. The dominant detection noise in the system typically is either detector dark noise for low detected power or optical shot noise for higher detected power [see Supplementary Materials], while at lower frequencies the laser frequency and amplitude technical noises may become significant as well, depending on the optical power and the instrument choice.

The Lorentzian fits (green lines) to the driven mechanical signal show the fundamental in-plane mode with a resonance frequency of ≈ 5.46 MHz and a Q of ≈ 31.4, and a much weaker signal originating from the out-of-plane mode at ≈ 1.35 MHz of a ≈ 9.8 quality factor. The AFM probe is designed to be used such that the measured physics quantities, such as topography and thermal properties of the substrate, generate mostly the in-plane motion[14,15] while the tip is in contact with a sample surface oriented normally to the chip plane[15]. Therefore, here we will only focus on the in-plane mode.

In the linear regime in Figure 4(a), the noise from thermal-mechanical fluctuations is largely below the detection noise, except for the frequency around the fundamental in-plane mode resonance. Because of the higher transduction gain with higher optical power in Figure 4(b) the thermal noise is amplified above the detector background over most of the frequency range. However, for both the driven motion signal and the thermal noise the increase of transduction gain is suppressed at low frequencies by the attenuation factor $r$, resulting in the gain saturation. The readout data are quantitatively described by the transfer function considering the thermo-optical nonlinearity (yellow line) where $r \approx 0.05$ is independently obtained from the optical spectrum and $\tau \approx 7.0$ μs from the numerical simulation is used.

Using the equipartition theorem and the measured thermal noise data, we calibrate the sensitivity to the in-plane displacement as ≈ 0.021 V/nm and ≈ 0.36 V/nm (at high frequency) for the linear and nonlinear cases, respectively [see Supplementary Materials]. We exclude the out-plane signal by subtracting the Lorentzian fit of the out-of-plane mode signal. The readout gain in the linear regime is frequency-independent, and the measured voltage is directly converted to displacement shown in Fig. 4(c). The frequency-dependent gain in the nonlinear regime is given by the transfer function, and the displacement is shown in Fig. 4(d). The increased transduction gain effectively improves the noise level of the optomechanical sensor from ≈ 6 fm/Hz$^{1/2}$ to ≈ 0.7 fm/Hz$^{1/2}$ at frequencies out of the resonance. In actual contact mode operation, the probe tip is touching the sample surface and the thermal noise is strongly suppressed[14,15]. Considering the detector background, the probe displacement is measured here with the uncertainty of ≈ 0.4 fm/Hz$^{1/2}$. Fig. 4(e) shows the improvement of the signal to noise ratio (SNR), achieved by increasing the optical power, as a function of the frequency. The gray data shows the improvement in performance for sensing the *displacement* of the probe, such as sample topography, while the red data is illustrating the performance improvement for sensing *force*, where the motion fluctuations driven by the thermal Langevin force impose an additional limit. In the case of sensing force, the increase of the transduction gain in the nonlinear regime improves the performance of the sensor from detection-noise-limited to thermal-noise-limited, and the improvement ratio depends on the ratio of the detection noise and the thermal noise in the linear regime. For example, the SNR improvement ratio at the eigenfrequency of the fundamental in-plane mode is around 1 since the thermal response is already much larger than the detection noise at the resonance in the linear regime [Fig. 4(a)], while it is between 1 and 10 at the frequency out of resonance. Another pronounced feature is that the improvement is lower than 1 at $f < 30$ kHz. It is attributed to the saturation of the transfer function at low frequency and the increasing detection noise with the increase of input optical power due to both shot and technical noises. As the gain increases linearly and shot noise increases as the square root of power, the displacement SNR always improves with power for high mechanical frequencies.

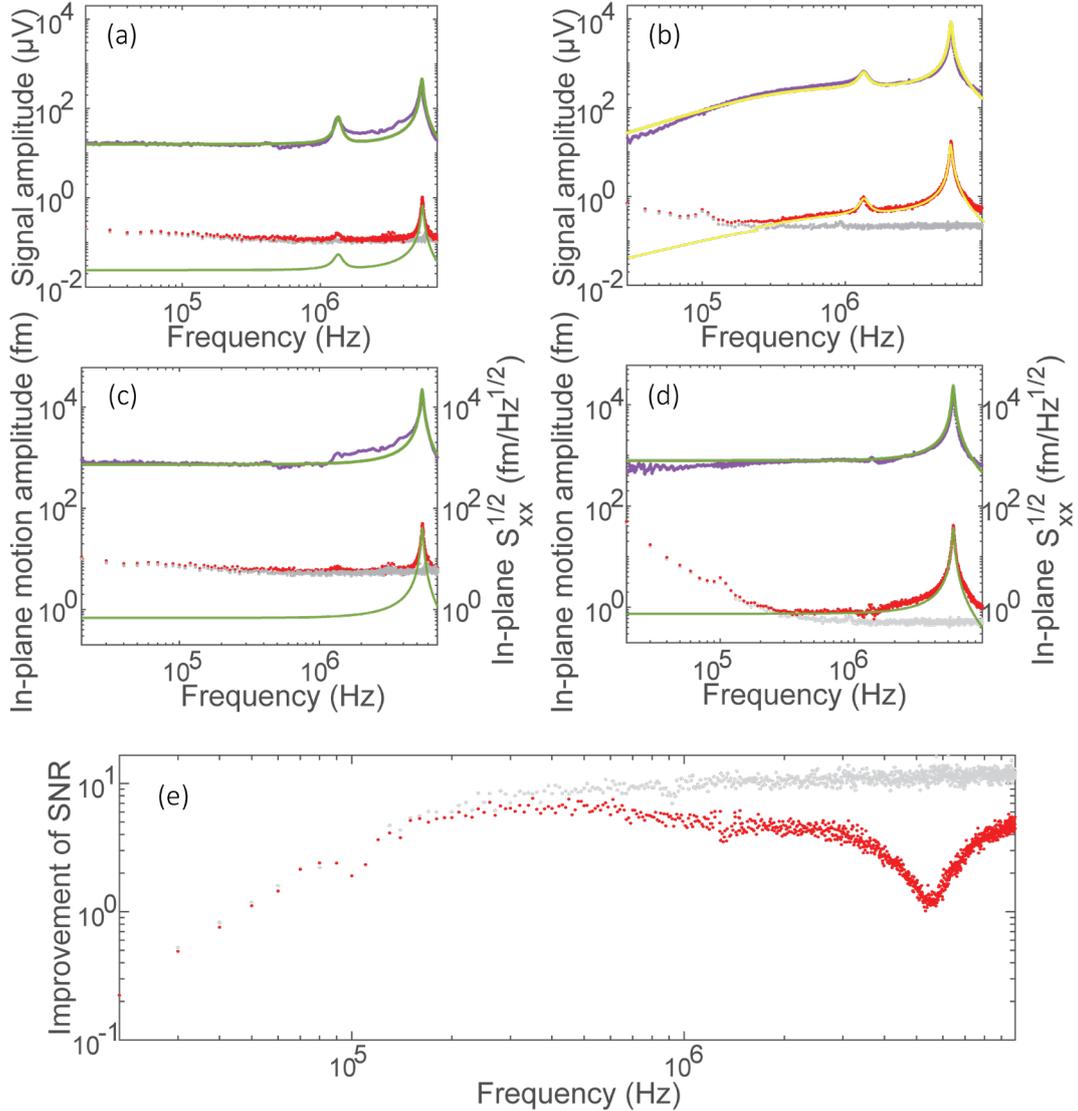

*Figure 4*. Sensitivity improvement in the thermo-optical regime. Electrostatically-driven response (purple), thermally-driven response plus detection noise (red), and solo detection noise (gray) are measured in the linear and nonlinear regime, shown in (a) and (b), respectively. Noise amplitudes are within 1 Hz equivalent bandwidth. Green lines are the corresponding Lorentzian fit. Yellow lines are Lorentzians multiplied by the transfer function, accounting for the thermo-optical nonlinearity. (c), (d) are calibrated in-plane motion amplitude from (a), (b) where the out-of-plane signals are removed. (e) Signal to noise ratio improvement for the force sensing limited by the thermal noise (red) and displacement sensing limited by detection noise (gray).

**Improving sensitivity of nanophotonic sensing by lowering optical quality factor**

Conventionally, the low-loss optical cavity with high Q is always desired for nanophotonic sensing since increasing the photon lifetime leads to longer interaction, higher transduction gain and increased sensitivity. Additionally, as we present above, the sensitivity can be further improved by increasing optical power. However, the combination of the high optical Q and high optical power does not guarantee the best sensitivity as intuitively expected, i.e. *an optical cavity with lower Q can have larger optimal (saturated) transduction gains and better sensing performance.* Qualitatively, this counterintuitive conclusion is due to the fact that for the low frequency input signals, the gains of high-Q optical modes saturate more quickly with increasing input power, while the gains of low-Q modes saturate later and possibly at higher gain values.

Figure 5(a) and (b) present two measured microdisk's WGMs with very different $Q \approx 10.0 \times 10^4$ and $\approx 1.4 \times 10^4$, respectively. When the input power is low, the blue lines show that the absolute value of the slope $k_{cold}$ of high-Q mode is higher, meaning a higher transduction gain, given an equal $g_{om,\lambda}$. For the high input power, the purple lines show that the high-Q mode in Fig.5 (a) is fully nonlinear (gain is saturated), while the lossy mode in (b) is still nearly linear (gain is not saturated) under the same input power and it has a higher $k_{hot}$ at the working wavelength despite its lower Q. In other words, the optimal (saturated) transduction gain of low-Q modes is shown to be better than that of the high-Q modes at low frequency.

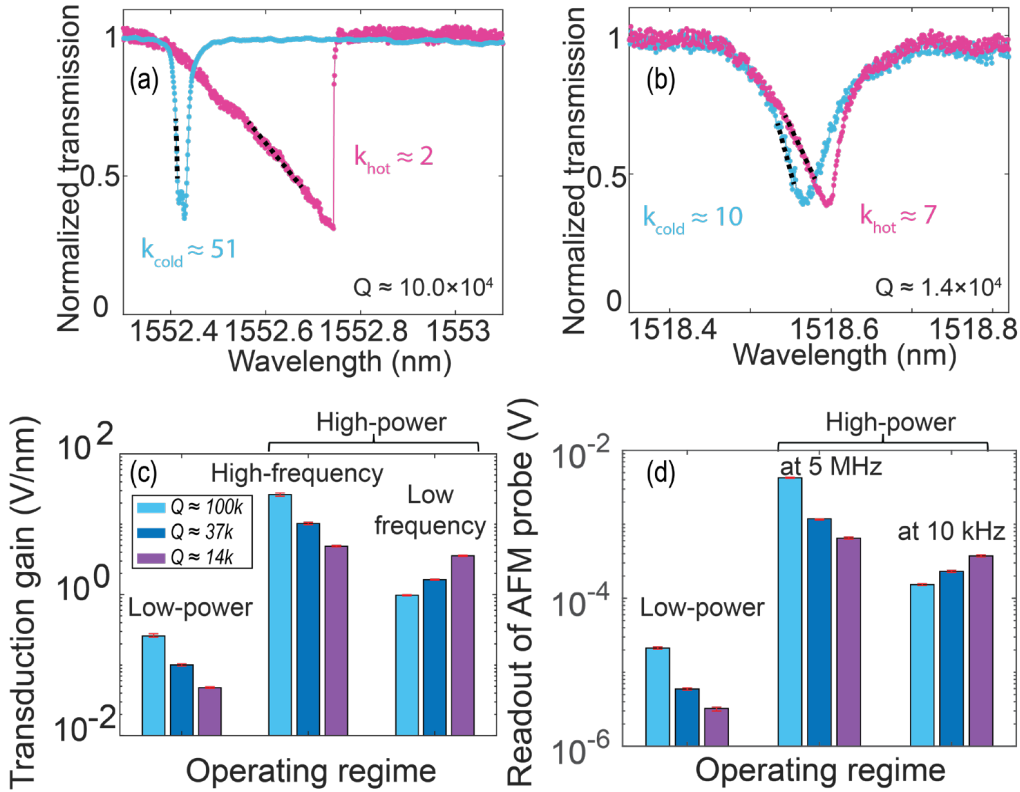

Figure 5. Transduction gain of nanophotonic modes with different quality factors. (a) and (b) are two measured optical modes from the first- and third-radial-order family of the microdisk, with very different Q $\approx 10.0 \times 10^4$ and $\approx 1.4 \times 10^4$, respectively. The blue line and purple line correspond to $I_i \approx 0.01$ V and $I_i \approx 0.98$ V, respectively. Their slopes at the working wavelength are label by the black dashed lines. (c) Transduction gains normalized by the $g_{om,\lambda}$. The light blue, dark blue, and purple bars corresponding to modes of radial order from 1 to 3, respectively. The operating

regimes from left to right correspond to lower-input-power regime ($I_i \approx 0.01$ V), high-input-power ($I_i \approx 0.98$ V) in high-frequency regime, and high-input-power in low-frequency regime where the transduction gain saturates due to thermo-optical tuning. The low-frequency gain at saturation counterintuitively increases with the decrease of Q. (d) Readout of the AFM probe at the corresponding regimes shown in (c). Ons standard deviation statistical uncertainties marked in (c) and (d) are obtained from the uncertaintes of the linear fit for the transmission dip, and the measured electrostatically-driven response at the specific frequencies, respectively.

The optomechanical transduction gains normalized by $g_{om,\lambda}$ are shown in Fig. 5(c), as determined by the slopes of the optical resonances. The modes of Q $\approx 10.0 \times 10^4, 3.7 \times 10^4$ and $1.4 \times 10^4$ are from the first- [Fig. 5(a)], second- [Fig. 2], and third-radial-order [Fig. 5(b)] families, and they are colored in light blue, dark blue, and purple, from left to right. At low input powers where the modes are linear and $\Delta T \approx 0$, the transduction gain $\frac{\Delta I}{\Delta x} = I_i g_{om,\lambda} k_{cold}$ based on Eq. (2). Therefore, modes with higher Q provide greater $k_{cold}$ and transduction gain, given equal values of $g_{om,\lambda}$. For high input powers, at high-frequency [$f \gg 1/(2\pi\tau)$], the transfer function is not affected by thermal dynamics, therefore, the transduction gain has the same trend as the low-power regime. By contrast, thermal dynamics at low-frequency results in saturating gain of the high-Q mode. The saturated gain is proportional to $rI_i = \frac{1}{\lambda_0 \alpha k_{cold} \eta}$, where $\lambda_0$ and $\alpha$ are constant for the modes under consideration. $\eta$ is proportional to the ratio of the energy absorption loss to the total energy loss in the cavity, given by the ratio of their loss rates: $\eta \propto \frac{\gamma_a}{\gamma_{total}}$, while $k_{cold} \propto \frac{1}{\gamma_{total}}$. Therefore the saturated gain at low frequency is $\propto \frac{\gamma_{total}^2}{\gamma_a}$. For nanophotonic resonators with multiple well confined modes, the absorption is mostly a cavity material property and tends to vary little with the mode order[24,25]. Large differences in Q and $\gamma_{total}$ arise largely from the different surface-scattering loss and radiative loss which strongly depend on the mode field distribution. Specifically for the nanophotonic cavity used in this paper, our recent dissipation analysis[24] shows $\gamma_a$ are nearly the same for the multiple studied modes (10 modes with 3 different radial orders) despite the order-of-magnitude difference in $\gamma_{total}$. As $\gamma_a$ is approximately constant, the saturated (maximum achievable) gain at low frequency increases rapidly with the increasing total dissipation rate, or decreasing Q.

We experimentally demonstrate the anomalous Q-dependence of the transduction signal on the AFM probe, shown in Figure 5(d). We drive the cantilever of the AFM probe to generate a fixed mechanical response, and measure the response using the three optical modes considered in Fig. 5(c). The readout signal for the same mechanical response at high frequency ($\approx 5$ MHz) increases with the Q of the optical transduction modes. However, the readout at low frequency ($\approx 10$ kHz) is the opposite, i.e. at the low frequency, the low-Q optical mode has higher optimal transduction gain than that of the high-Q mode.

## Conclusion and outlook

In summary, we derived and experimentally validated a broadband, small-signal transfer function valid for the high optical power, thermo-optically nonlinear regime of photonic sensing.

The transfer function uses the thermal time constant independently obtained from finite element modeling and the gain ratios separately measured from optical spectra. Excellent agreement with the optomechanically transduced power spectral densities of a nanocantilever is obtained without adjustable parameters, enabling quantitative optomechanical motion transduction with high sensitivity over a large frequency bandwidth down to DC. The displacement measurement noise of our optomechanical AFM probe improves from the best achievable performance $\approx 6 \text{ fm/Hz}^{1/2}$ in the linear regime to $\approx 0.4 \text{ fm/Hz}^{1/2}$ in the thermo-optical nonlinear regime. Not limited to optomechanical transducers, this work demonstrates a general and practical approach for improving the sensitivity of a broad range of technologically important photonic resonator sensors subject to thermo-optical tuning via optical absorption.

Additionally, the interplay between the thermo-optical effect and the dissipations of the nanophotonic cavity reveals a new counterintuitive regime where the low-Q modes possess higher optimal transduction gain at low frequency. The new regime may inspire new strategies of design for nanophotonic sensors for low-frequency and static applications, such as topography scanning[26] and sensing dynamics of cells[27].

## Materials and Methods

### Numerical simulation

The numerical simulation of the thermal dynamics of the microdisk is performed by commercial finite-element-method software. The geometries used in the simulation include all relevant components, including the microdisk, supporting post, cantilever, frame, and substrate. Since the supporting post is the main thermal impedance we carefully measured its dimension on a testing sample released together with the device. The diameter of the top and bottom surface of the post defined by the hydrofluoric acid etching process is measured to be ≈ 1.5 μm and ≈ 2.2 μm. The whole setup is in an air environment.

In the simulation, input power is applied at time zero to the rim of the microdisk where the optical modes are located. By fitting the thermal response of the microdisk, we obtain its thermal constant $\tau \approx 7.0$ μs [see Supplementary Materials].

Acknowledgments
We thank Dr. Jeffrey Schwartz, Dr. Biswarup Guha, Dr. J. Alexander Liddle and Dr. Marcelo Davanco for reviewing this paper and giving meaningful suggestions. M.W. and D.J.P. are supported by the Cooperative Research Agreement between the University of Maryland and the National Institute of Standards and Technology Center for Nanoscale Science and Technology, Award 70NANB14H209, through the University of Maryland.

Author contributions
V.A. conceived the project. M.W. and D.J.P designed the research. D.J.P. and M.W. fabricated the devices. M.W. and D.J.P. performed the experiments. M.W. analyzed the data. All authors discussed the results and contributed the manuscript.

Conflict of interest
The authors declare that they have no conflict of interest.

# Supplementary Materials

## Overcoming Thermo-Optical Dynamics in Broadband Nanophotonic Sensing

Mingkang Wang[1,2,†], Diego J. Perez-Morelo[1,2], Vladimir Aksyuk[1,*]

[1]Microsystems and Nanotechnology Division, National Institute of Standards and Technology, Gaithersburg, MD 20899 USA

[2]Institute for Research in Electronics and Applied Physics, University of Maryland, College Park, MD 20742, USA

1. **Numerical simulation of the thermal effect**

As described in the main text, our nominally 10 μm diameter, 260 nm thick silicon microdisk is supported at the center by a 2 μm thick silica post. The diameter of the top and bottom surface of the post defined by the hydrofluoric acid etching process is measured to be ≈ 1.5 μm and ≈ 2.2 μm. The whole setup is in an air environment.

In the simulation, an input power of $P_{in}$ ≈ 0.1 mW is applied at time zero to the rim of the microdisk where the optical modes are located. Figure S1 (a) shows the temperature distribution of the system at the thermal equilibrium. The post is the main source of thermal impedance, while the Si substrate has high-enough thermal conductivity to be nearly isothermal. The effect of the cantilever on the simulation results is less than 2%, however, for the accuracy, we also included the cantilever in the model. The time-domain thermal simulation result is presented in Figure S1 (b). By fitting the curve with $T = T' + \Delta T e^{-\frac{t}{\tau}}$, we obtain the thermal response time $\tau$ ≈ 7.0 μs, and κ ≈ $1.5 \times 10^5$ K/W. $T' = 293$ K is a constant offset.

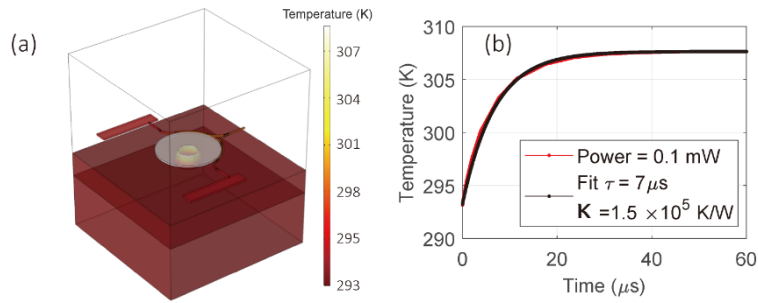

Figure S1. Simulated thermal response of the silicon microdisk. (a). Thermal distribution on the microdisk at equilibrium. (b). Dynamic response of the disk to the application of a known thermal load at the periphery.

## 2. Transduction gain saturation

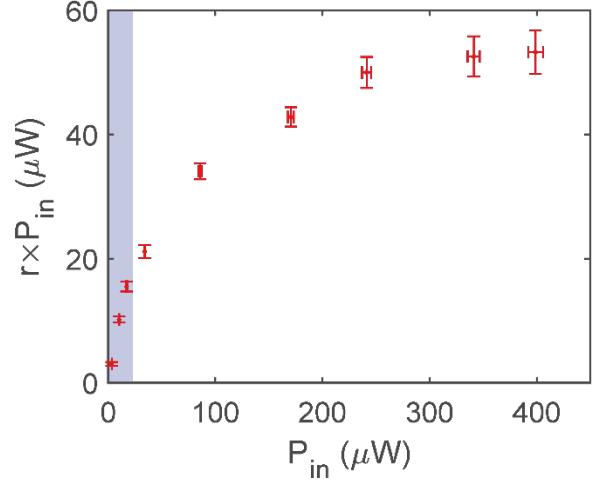

Figure S2. Product of the transfer function parameter $(r)$ and the input power $(P_{in})$ as a function of $P_{in}$. $rP_{in}$ approaches saturation as $r$ decreases with increasing $P_{in}$.

Figure S2 shows the measured $rP_{in}$ where $r = \frac{k_{hot}}{k_{cold}}$ is measured from the transmission signal at the corresponding $P_{in}$. It shows that the transduction gain, $\propto rP_{in}$, at zero-frequency tends to saturate for high input powers, becoming independent of the input power $P_{in}$.

Note, the saturation only occurs at zero-frequency in the nonlinear regime, the linear dependence of the gain on $P_{in}$ at high frequency is still valid.

## 3. Probe response noise in the frequency domain and probe response equipartition calibration

Figure S3 (a) and (b) show the measured noise power spectral densities of the AFM probe with a double-clamped cantilever in the nonlinear and linear regime, respectively. The dark noise (black) of the photodetector is measured when the input optical power is set to 0, the shot noise + dark noise is measured when the same optical power is transmitted at the working wavelength set at the background of the optical transmission spectrum (transduction gain = 0), and the thermal noise (red) is finally included when the working wavelength is tuned to the shoulder of the optical dip. It is noteworthy the total noise power spectral density (red) also contains the experiment-dependent technical noises such as optical fiber vibration and laser frequency instability, however, they mainly exist at low mechanical frequency range (<0.1 MHz).

We convert the voltage power spectral density to input-optical-power-dependent displacement noise [Fig. 4(a) (c)] by using the equipartition theorem. First, we modeled the device with a finite element method (FEM) which gives the correct mechanical eigenfrequencies, the same as the measuring device. Next, we obtain the effective spring constant of the fundamental in-plane mode as $k_{in} \approx 2.03$ N/m. Second, the integrated voltage noise power [area under the green line in Fig. S3(a)] is equal to $\langle x_{in}^2 \rangle = \frac{k_B T}{k_{in}}$ as the result of the equipartition theorem. It gives the calibration constant $a_1 \approx 2.8$ fm/μV for (a). As the transduction gain depends on the input power, the calibration constant also depends on the input optical power. However, as the calibrated displacement does not change with gain, we can easily obtain the calibration constant for (b) as $a_2 \approx 48.1$ fm/μV. $a_1$ and $a_2$ are used in the calibration of Fig. 4(c) and (a), respectively.

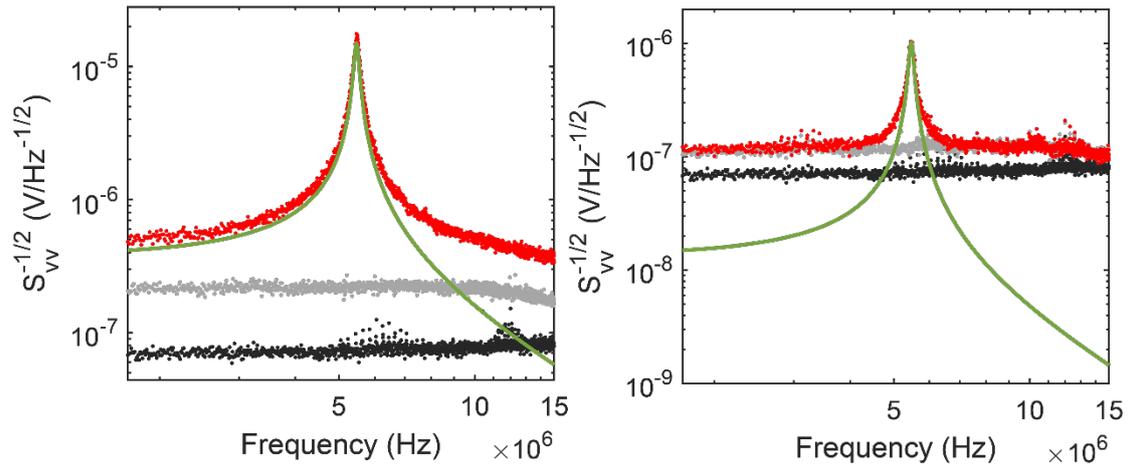

Figure S3. Power spectral density of dark noise (black), dark noise+shot noise (gray), dark noise+shot noise+thermal noise (red), and the Lorentzian fit to the thermal noise only (green) in nonlinear case (a) and linear case (b).